\documentclass[twocolumn,english,superscriptaddres,noshowpacs,amssymb,floatfix]{revtex4}

\setcitestyle{super}

\usepackage{graphicx}

\newcommand{\erf}[1]{Eq.~(\ref{#1})} 

\usepackage{amsmath}
\usepackage{braket}
\usepackage{color}

\def\urlprefix{}
\def\url#1{}

\begin{document}


\title{Dynamics of simultaneously measured non-commuting observables}


\author{Shay Hacohen-Gourgy$^{\dagger,1,2}$}
\email[shayhh@berkeley.edu]{shayhh@berkeley.edu}
\author{Leigh S. Martin$^{\dagger,1,2,3}$}
\author{Emmanuel Flurin$^{1,2}$}
\author{Vinay V. Ramasesh$^{1,2}$}
\author{K. Birgitta Whaley$^{3,4}$}
\author{Irfan Siddiqi$^{1,2}$}
\affiliation{$^1$Quantum Nanoelectronics Laboratory, Department of Physics, University of California, Berkeley CA 94720, USA.}
\affiliation{$^2$Center for Quantum Coherent Science, University of California, Berkeley CA 94720, USA.}
\affiliation{$^3$Berkeley Center for Quantum Information and Computation, Berkeley, California 94720, USA.}
\affiliation{$^4$Department of Chemistry, University of California, Berkeley, California 94720, USA.}

\maketitle

\textbf{
In quantum mechanics, measurements cause wavefunction collapse that yields precise outcomes, whereas for non-commuting observables such as position and momentum Heisenberg’s uncertainty principle limits the intrinsic precision of a state. Although theoretical work\cite{Arthurs1965} has demonstrated that it should be possible to perform simultaneous non-commuting measurements and has revealed the limits on measurement outcomes, only recently\cite{Jordan2005, Ruskov2010}  has the dynamics of the quantum state been discussed.  To realize this unexplored regime, we simultaneously apply two continuous quantum non-demolition probes of non-commuting observables to a superconducting qubit. We implement multiple readout channels by coupling the qubit to multiple modes of a cavity. To control the measurement observables, we implement a ‘single quadrature’ measurement by driving the qubit and applying cavity sidebands with a relative phase that sets the observable. Here, we use this approach to show that the uncertainty principle governs the dynamics of the wavefunction by enforcing a lower bound on the measurement-induced disturbance. Consequently, as we transition from measuring identical to measuring non-commuting observables, the dynamics make a smooth transition from standard wavefunction collapse to localized persistent diffusion and then to isotropic persistent diffusion. Although the evolution of the state differs markedly from that of a
conventional measurement, information about both non-commuting observables is extracted by keeping track of the time ordering of the measurement record, enabling quantum state tomography without alternating measurements. 
Our work creates novel capabilities for quantum control, including rapid state purification\cite{Ruskov2012}, adaptive measurement\cite{Jacobs2003project,Wiseman1995adaptive}, measurement-based state steering and continuous quantum error correction\cite{Ahn2002}. As physical systems often interact continuously with their environment via non-commuting degrees of freedom, 
our work offers a way\cite{Dressel2016LG,Nishizawa2015ContinuousErrorDisturbance} to study how notions of contemporary quantum foundations\cite{Kirchmair2009,Kochen1968,Erhart2012,DresselErrorDist,Ozawa2003} arise in such settings.
}

In this work, we implement two high quantum efficiency readouts of the angular momenta about two different axes of an artificial spin-1/2 system and observe in real-time the resulting dynamics. Importantly, our measurements are designed to be individually quantum non-demolition, which ensures that the back-action arises only from the competition between incompatible observables.

Our experiment utilizes a single superconducting transmon qubit coupled dispersively to a multimode waveguide cavity\cite{Koch2007}. This results in a qubit-state dependent shift of the cavity mode frequency. 
By applying a microwave tone to a single cavity mode, one can infer the qubit state from the phase of the reflected signal\cite{Gambetta2008};
this readout scheme has been used extensively with superconducting qubits for quantum information processing, and also to perform weak measurements\cite{Hatridge2013} and track quantum trajectories of a single qubit\cite{Murch2013}. In our configuration, each cavity mode constitutes a measurement channel which extracts the projection of the qubit spin along the $\sigma_z$ axis of the Bloch sphere. 

\begin{figure}[h!]
\begin{center}
{\includegraphics[width=\linewidth, bb=0 0 350 400]{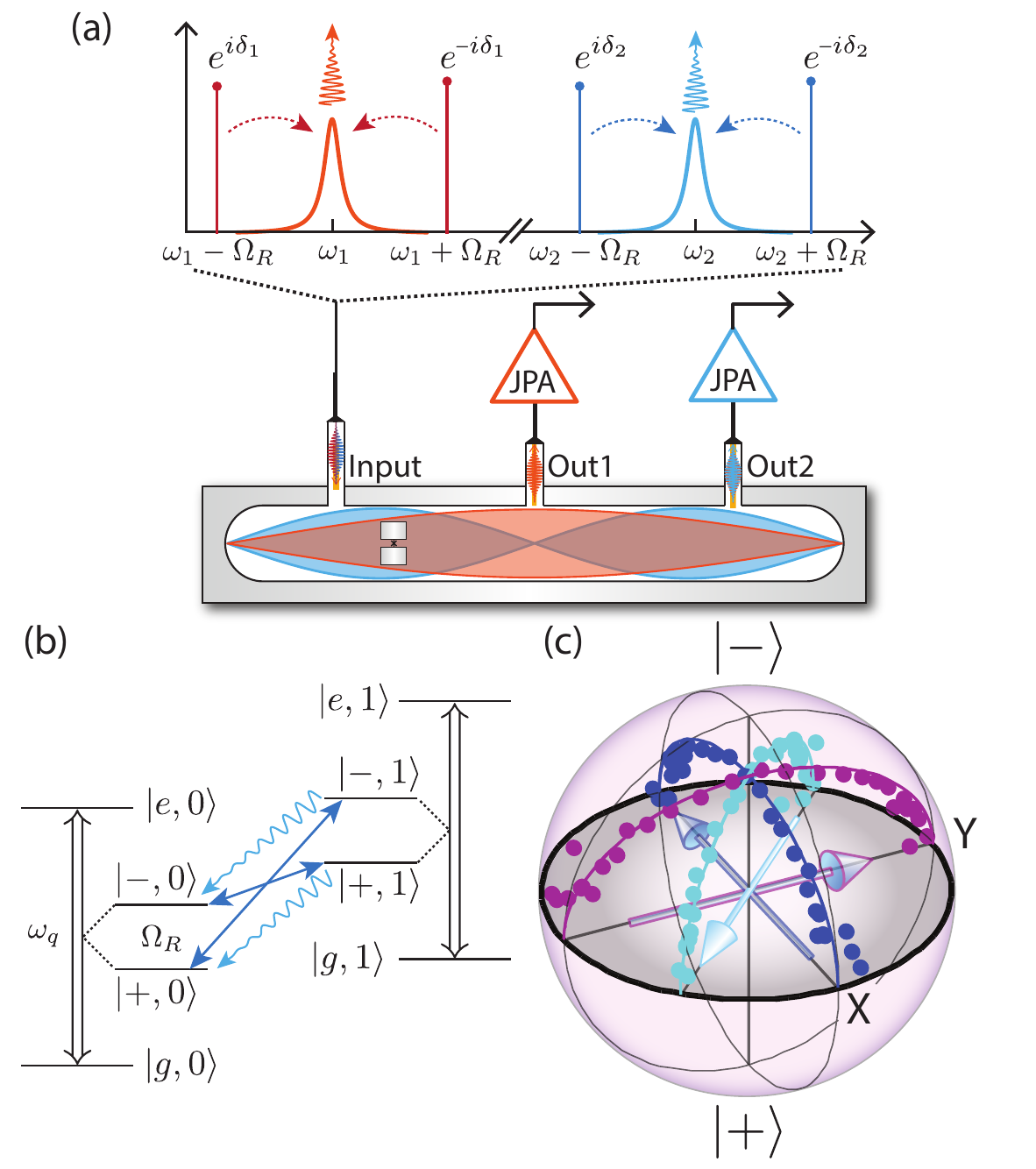}}
\caption{\textbf{Multimode single quadrature measurement (SQM)} (a) A transmon qubit in an aluminum cavity. Via the input port, we drive Rabi oscillations at frequency $\Omega_R$ and apply $\pm\Omega_R$ sideband tones to each of the two lowest cavity modes. Each output is monitored using a lumped-element Josephson parametric amplifier (LJPA)\cite{Beltran2008Sup}.
Coupling is designed so that $\sim$90\% of each signal is routed to its corresponding LJPA. This system yields two separate measurement channels, each with a controllable measurement basis. (b) Dressed state picture of the SQM scheme for one of the cavity modes. Sideband tones drive transitions indicated by the solid lines, and undulating lines represent cavity decay, which we detect. (c) Tomographic reconstruction after a $1~\mu \mathrm{s}$ SQM applied only to the lower mode, showing collapse along three separately chosen measurement axes $\sigma_{\delta_1}$, $\delta_1 = \{ 0,\pi/4,\pi/2 \} $. Circles plot tomographic data and lines give theory based on a dephasing rate of $\Gamma_{1}/2 \pi = 122 ~\mathrm{kHz}$ and quantum efficiency of $\eta_1 = 0.41$.}
\label{fig:Setup}
\end{center}
\end{figure}

The key ingredient to measuring non-commuting observables is to re-engineer the underlying Hamiltonian so that each cavity mode couples to a controllable axis of the qubit. We accomplish this with what we call a single quadrature measurement (SQM), motivated by the back-action evading techniques recently conceived for optomechanical systems\cite{Caves1980,Hertzberg2010}. The central idea is to drive Rabi oscillations $\Omega_R/2\pi = 40 ~\mathrm{MHz}$ on the qubit so that its Hamiltonian becomes that of an effective low frequency qubit. As depicted in Fig.\ref{fig:Setup}a, we apply a pair of sideband tones detuned above and below the cavity frequency by $\Omega_R$, which the qubit then Raman scatters to the cavity resonance. Separately, the 
red(blue)-detuned sidebands
would induce cavity-mediated cooling(heating) of the effective qubit\cite{Murch2012}, as illustrated in Fig. \ref{fig:Setup}b. When they are applied concurrently, 
their relative phase determines the interference between the up- and down-converted photons, which in turn dictates which qubit observable is encoded in the resulting signal.
The dynamics are governed by a coherent sum of the two effective Hamiltonians

\begin{align}
\label{eq:Heff}
H_\text{eff} &= 
\frac{\chi \bar{a}_0}{2} \Big[ \underbrace{a \sigma^\dagger e^{i \delta} + a^\dagger \sigma e^{-i \delta}}_\text{resonant heating} + \underbrace{a \sigma e^{-i \delta} + a^\dagger \sigma^\dagger e^{i \delta}}_\text{resonant cooling} \Big] \nonumber \\
&= \frac{\chi \bar{a}_0}{2} (a+a^\dagger) \sigma_{\delta}
\end{align}

where $\chi$ is the qubit-cavity coupling, $\bar{a}_0$ is the sideband amplitude, $a$, $a^\dagger$ are the cavity ladder operators and $\sigma$, $\sigma^\dagger$ are the raising and lowering operators with respect to the dressed basis $\ket{\pm}=(\ket{g}\pm\ket{e})/\sqrt{2}$. Note that this scheme implements longitudinal coupling, an important feature for squeezing-enhanced readout\cite{Didier2015}. Importantly, the relative sideband phase $\delta$ sets the coupling axis $\sigma_\delta \equiv \sigma_x \cos \delta +\sigma_y \sin\delta$. The resulting Hamiltonian is a resonant cavity drive, the sign of which depends on the qubit state along the $\sigma_\delta$ axis. Detecting the cavity output field with quantum efficiency $\eta$ yields a measurement of the qubit at a rate $\Gamma \eta = 2 \chi^2 \bar{a}_0^2 \eta/\kappa$ in the $\sigma_\delta$ basis\cite{Gambetta2008} (see methods), which we can now control. 

\begin{figure}
{\includegraphics[width=\linewidth]{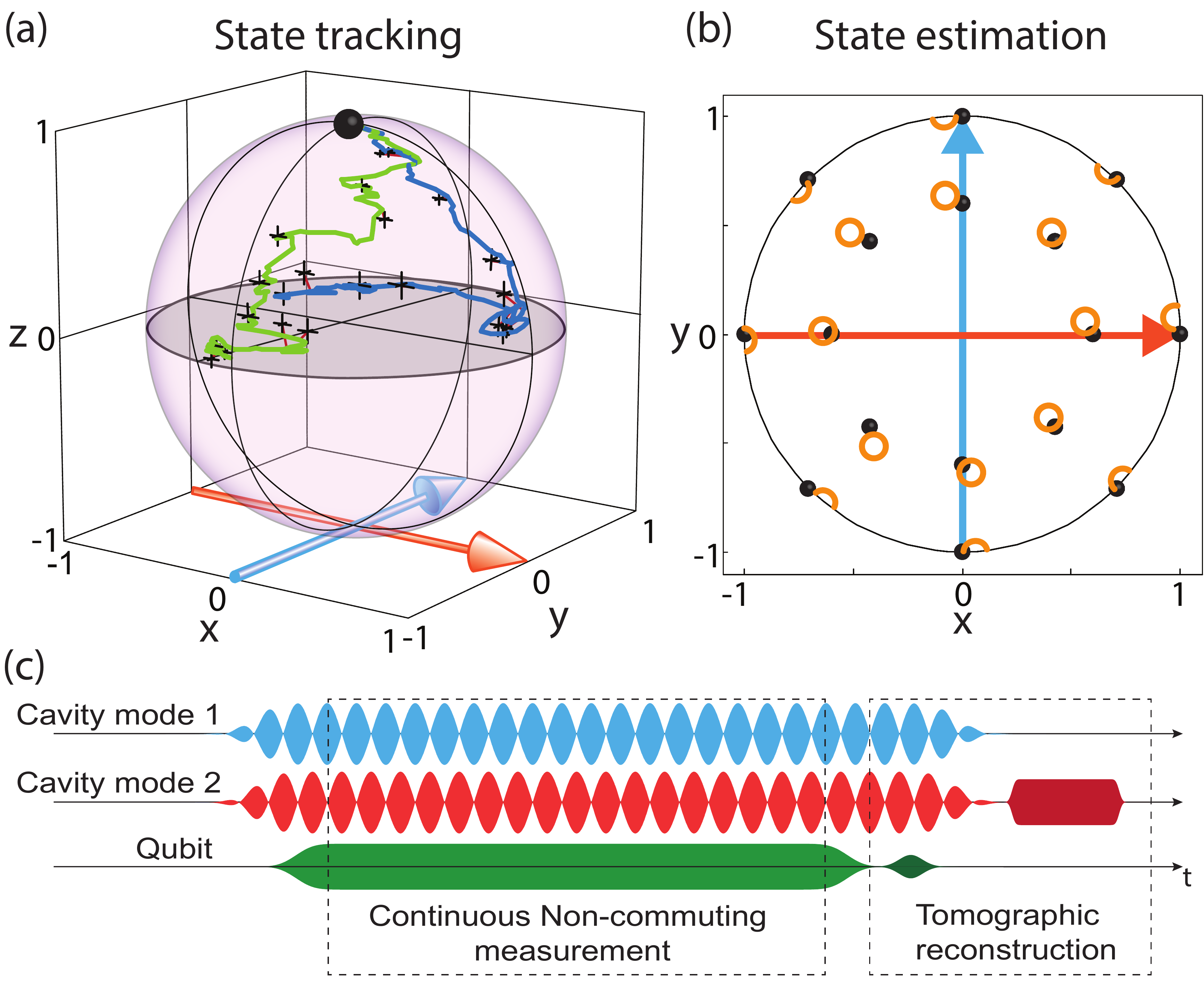}}
\caption{\textbf{Validation of simultaneous non-commuting measurements} (a) Reconstruction of two quantum trajectories initialized at state (green and blue lines) $\ket{-}$ (black sphere) and their associated tomographic validation (black crosses). Red and blue arrows indicate axis of each measurement. Tomography points are taken every $200~\text{ns}$, and are each linked to the corresponding trajectory time point with a maroon line. All trajectories ending within $\pm 0.11$ around a given point along the plotted trajectory are used for the corresponding tomography reconstruction. Error bars are generated from statistical uncertainty arising in qubit readout. (b) Estimation of 16 initially unknown state preparations (black spheres) using non-commuting measurements. Orange circles mark 95\% confidence intervals, generated by applying maximum likelihood estimation to $\sim$10,000 trajectories each. (c) Pulse sequence, showing Rabi drive (green) and sideband tones applied to cavity modes 1 and 2 (blue and red respectively), which generate measurements indicated by arrows in panels a and b. Tomographic reconstruction consists of a tomography qubit pulse and projective readout.}
\label{fig:Tomography}
\end{figure}

\begin{figure*}
{\includegraphics[width=1.4\linewidth, bb=0 0 750 250]{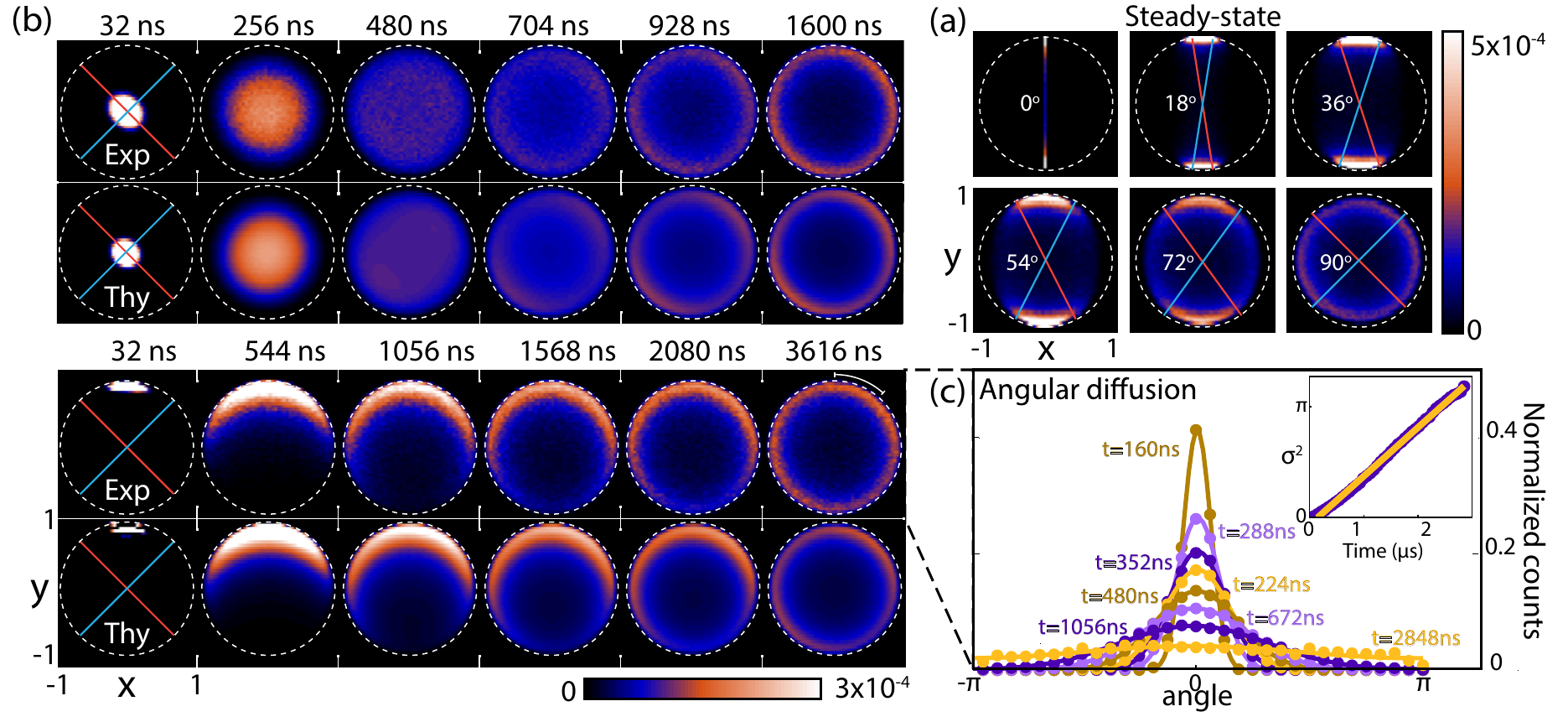}}
\caption{\textbf{Probability distribution of the density matrix} (a) in the steady-state as function of angle, between the axes, demonstrating loss of collapse for non-commuting observables. (b) as function of time for perpendicular measurement axes. We prepare a mixed state for observing the radial dynamics (top), and a state with purity of $P = 0.89$ for observing the azimuthal dynamics (bottom). Upper row of each is experimental data, and bottom row is theoretical comparison derived from the Fokker-Planck equation (see methods). (c)
Angular probability distributions for states within a ring of inner radius 0.86 and outer radius 0.92, showing that dynamics match those of a random walk. Points are normalized counts and lines are fits to normal distributions convolved with a $2 \pi$ periodic Dirac comb. Inset -- variance as a function of time (violet) and linear fit (yellow) yielding a slope of 1.4 $\mu \mathrm{s^{-1}}$. The expected slope for a perfect random walk in our system is 1.5 $\mu \mathrm{s^{-1}}$ with the main source of uncertainty due to measurement rate uncertainty of about 10\%.
}
\label{fig:FokPlanck}
\end{figure*}

\begin{figure}
{\includegraphics[width=\linewidth, bb=0 0 350 250]{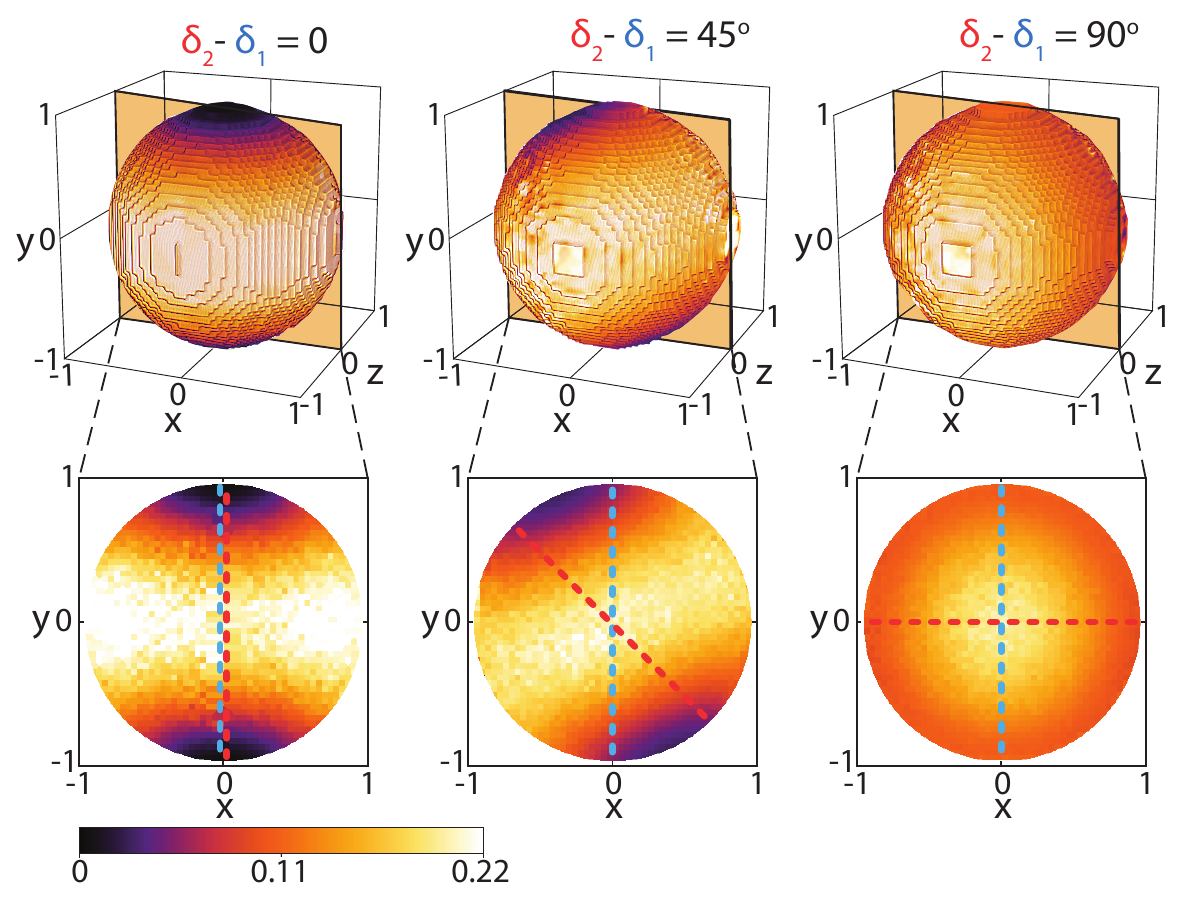}}
\caption{\textbf{Magnitude of the quantum back-action.} The amplitude of disturbance induced by measurement ($\text{Tr}[d\rho^\dagger d\rho]$) at each point on the Bloch sphere for three relative measurement angles $\delta_2-\delta_1$. (top) Disturbance at the surface of the Bloch sphere, which is bounded from below by the uncertainty principle. (bottom) A slice through the $z=0$ plane. Dotted lines mark measurement axes. For non-commuting observables ($\delta_2-\delta_1 \neq 0$), no point of zero disturbance exists within the entire volume of the Bloch sphere, indicating that the state must diffuse indefinitely.}
\label{fig:Name4}
\end{figure}

First, to demonstrate control of the measurement axis, we perform a single SQM using the lower cavity mode. We prepare the $|-\rangle$ state and read out for $1~\mu\text{s}$ followed by a tomography rotation and a projective measurement using a standard dispersive readout of $\sigma_z$ in the lab frame. During the SQM, drift in the Rabi frequency and spurious tones at the cavity frequency induce unwanted rotations about the $\sigma_z$ and $\sigma_\delta$ axes respectively (see methods). To mitigate these effects, we actively feedback to stabilize the Rabi frequency to within $10~\text{kHz}$ and suppress leakage of the local oscillator used to generate the sidebands to the $10^{-4}$ photon level. Because a single SQM commutes with itself at all times, the integral of the signal fully specifies the amount of projection onto an eigenstate\cite{Murch2013}. We perform tomographic state validation on subsets of the data post-selected according to this integrated signal. Tomography data in Fig. \ref{fig:Setup}c show state collapse along the desired axis, confirming control of the measurement operator $\sigma_\delta$.

Next, to simultaneously measure two non-commuting observables, we apply the SQM to the two lowest cavity modes. Each pair of sidebands sets a measurement axis, and thus we are able to control them independently. Because non-commutativity leads to dependence on the time ordering of measurement outcomes, the integral of the output signal is no longer sufficient to uniquely determine the qubit state in general. Nevertheless, the combined measurement records encode complete information, and we are able to track quantum trajectories of an initially known state. We calculate the trajectories by iteratively applying a measurement operator which corresponds to the record of both $\sigma_{\delta_1}$ and $\sigma_{\delta_2}$ integrated over a short interval $\Delta t = 16~\mathrm{ns}$ (see methods).
To experimentally verify the trajectories, we perform quantum state tomography\cite{Nielson2010} on subsets of trajectories that end at approximately the same point on the Bloch sphere. Figure \ref{fig:Tomography}a shows two such traces with their tomographic validation from data in which the measurement axes are set to be orthogonal, $\sigma_{\delta_1}=\sigma_x$ and $\sigma_{\delta_2}=\sigma_y$. 

Due to the disturbances introduced by measurement incompatibility, extraction of an initially unknown quantity, such as a Hamiltonian parameter or system observable, requires use of the \textit{combined} measurement records and their full time orderings.
In particular, estimation methods must rely not only on the statistics of the measurement records, but also on some estimate of this disturbance. We encode this information in a set of composite measurement operators, one for each trajectory\cite{Six2016} (see methods), and use them to reconstruct a set of initial state preparations, analogous to state tomography.
We then perform maximum likelihood estimation of the initial state over a large set of trajectories taken with the same initial state\cite{Six2016}. Figure \ref{fig:Tomography}b shows reconstruction of sixteen state preparations again in the case of orthogonal measurement axes. Agreement within the confidence interval demonstrates that despite the complicated dynamics induced by competing observables, our scheme performs as a measurement, and that it extracts information about $\sigma_x$ and $\sigma_y$ simultaneously. 

A state undergoing a non-commuting measurement exhibits dynamics beyond those of usual wavefunction collapse. We directly observe this evolution by measuring the probability distributions of the resulting density matrices. Figure~\ref{fig:FokPlanck}a shows the steady state probability distributions. When the axes align, the system collapses to one of two points at the poles of the Bloch sphere as expected for commuting measurement operators. When the axes are separated by a finite angle less than 90 degrees, the state does not collapse to any point, but rather 
localizes to regions of finite size defined by the measurement axes.
A salient feature is that when the axes are orthogonal and hence maximally incompatible, the location of the measurement axes no longer leaves any imprint on the state evolution. The distinct regions merge and we lose all notion of collapse.

Figure~\ref{fig:FokPlanck}b shows probability distributions as a function of time for this canonical case. Starting in a mixed state, we see that the state purifies isotropically to a mean steady-state radius given by $r = \sqrt{\eta}$. 
High quantum efficiency in our system allows us to observe the azimuthal dynamics by preparing a state of purity $P = 0.89$, the most likely steady-state purity. As predicted by Ruskov \textit{et al.} \cite{Ruskov2010} in the case of unit quantum efficiency, competition between the maximally incompatible observables leads to diffusive motion given by a uniform random walk. Figure \ref{fig:FokPlanck}c shows the angular distribution as a function of time, which agrees quantitatively with this prediction.

The diffusive behavior seen in Fig. \ref{fig:FokPlanck}c suggests that even once the probability distributions have reached steady state, the system continues to evolve. To quantify this measurement-induced disturbance, we plot the norm of the state change $d\rho$ over an interval of $64~\mathrm{ns}$ versus position on the Bloch sphere in Fig. \ref{fig:Name4}. When the measurements are compatible, two points on the Bloch sphere exhibit zero disturbance, which indicate mutual eigenstates of the measurements and hence points of collapse. No such points exist when the measurements are incompatible, implying that the state diffuses indefinitely. If we calculate this disturbance directly from the stochastic master equation for the system, we find the following relation valid for all initially pure states

\begin{align}
\label{eq:distmap}
\text{Tr}[d\rho^\dagger d\rho] = (\Delta \sigma_{\delta,1}^2 \Gamma_{1}\eta_1 + \Delta \sigma_{\delta,2}^2 \Gamma_{2}\eta_2) dt \\ \nonumber
\geq |\langle[\sigma_{\delta_1}, \sigma_{\delta_2}]\rangle|\sqrt{\eta_1 \eta_2 \Gamma_{1}  \Gamma_{2}} dt
\end{align}

which holds for any Hermitian operators of any system. The right hand side of the equality closely resembles the original Heisenberg uncertainty relation, but contains the sum of the variances instead of the product. Unlike the latter, the sum can be bounded by a stronger inequality which is never trivial for non-commuting measurements\cite{MacCone2014}. 
This shows that the persistent diffusion observed in Fig. \ref{fig:FokPlanck}c is a universal consequence of the uncertainty principle and can be quantitatively derived from it.

The techniques we developed readily generalize to multi-level and multi-qubit systems, and have numerous potential applications for quantum information science. The ability to measure multiple distinct operators of a quantum system simultaneously could allow for implementation of continuous quantum error correction, in which several error syndromes can be monitored concurrently\cite{Ahn2002}. Furthermore it may motivate development of quantum-limited metrology protocols that acquire information simultaneously about multiple incompatible properties. By monitoring a third cavity mode and applying an additional sideband drive, the axis perpendicular to the Rabi plane can also be probed\cite{Vool2016}. With a modest improvement in quantum efficiency, our scheme will also lead to faster initialization of quantum circuits\cite{Ruskov2012, Combes2010}.

Our experiment reveals the subtle interplay between the Heisenberg uncertainty principle and wavefunction collapse. As incompatibility of observables is at the core of quantum theory\cite{DresselPhD}, tests of quantum foundations must access these properties.
Existing work on quantum foundations has focused on testing the validity of essential features of quantum mechanics, such as contextuality and the various bounds on error and disturbance. Our work presents the possibility of exploring how such concepts emerge in realistic systems\cite{Nishizawa2015ContinuousErrorDisturbance, Dressel2016LG}, which tend to interact continuously with their environments via multiple non-commuting decoherence channels. Thus our work provides an indispensable tool for investigating this virtually unexplored territory of quantum foundations.\\

$^\dagger$ These authors contributed equally to this work.

\textbf{Acknowledgements} We thank Justin Dressel, Andrew Jordan, Alexander Korotkov, Joshua Combes, Mohan Sarovar, Uri Vool and Juan Atalaya for invaluable discussions, and MIT Lincoln labs for fabrication of the TWPA. L.S.M. and V.V.R. acknowledge support from the National Science Foundation Graduate Fellowship Grant No. 1106400. L.S.M. additionally acknowledges support from the Berkeley Fellowship for Graduate Study. This work was supported by the Air Force Office of Scientific Research under Grant \# FA9550-12-1- 0378 and the Army Research Office under Grant \# W911NF-15-1-0496.

\textbf{Author Contributions} S.H. and L.S.M. wrote the manuscript, conducted the experiment and analyzed the data. S.H., L.S.M., E.F. and I.S. conceived of the experiment. S.H. and L.S.M. constructed the experimental setup with assistance from V.V.R. and E.F.. S.H. fabricated the qubit and LJPAs. L.S.M. performed theoretical analysis with assistance from S.H. All authors contributed to discussions and preparation of the manuscript. All work was carried out under the supervision of K.B.W. and I.S..

\textbf{Author Information} The authors declare no competing financial interests.

\begin{center}
\includegraphics[width = 50mm, bb=0 0 700 300]{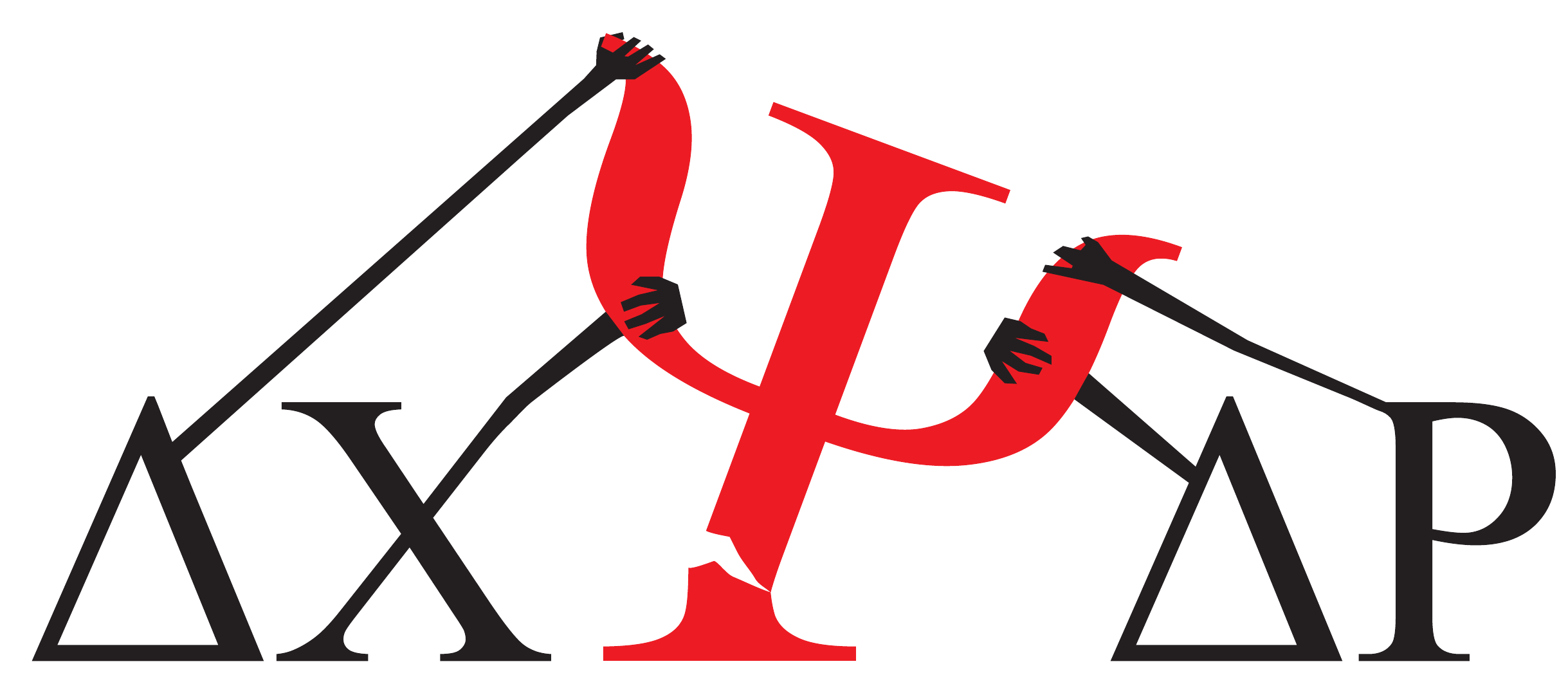}
\end{center}

\section{Methods}
\textbf{Experimental setup and sample.} Our transmon qubit is fabricated on double-side-polished silicon, with a single double-angle-evaporated Al/AlOx/Al Josephson junction. The internal dimensions of the aluminum 3D cavity are 81 mm x 51 mm x 20 mm. The qubit is positioned 23 mm from the edge of the cavity. The qubit is characterized by a charging energy $E_c/h$ = 220 MHz and a $\ket{0}$ to $\ket{1}$ transition frequency $\omega_q/2\pi$ = 4.262 GHz. The qubit coherence is characterized by an excited state lifetime of $T_1 = 60~\mu\mathrm{s}$, echo time of $T_{2,\text{echo}} = 40~\mu\mathrm{s}$ and Rabi decay time of $25 ~\mu\mathrm{s}$. The two lowest cavity modes used in the experiment have frequencies $\omega_1/2 \pi$ = 6.666 GHz, $\omega_2/2 \pi$ = 7.391 GHz, linewidths $\kappa_1/2 \pi$ = 7.2 MHz, $\kappa_2/2 \pi$ = 4.3 MHz and qubit dispersive frequency shift $\chi_1/2 \pi$ = 0.18 MHz, $\chi_2/2 \pi$ = 0.23 MHz respectively. The cavity outputs are amplified using two lumped-element Josephson parametric amplifiers (LJPA) operated in phase sensitive mode. Amplifier gains are set to 15 dB for mode 1 and 18 dB for mode 2. To prevent saturation of the amplifiers by the SQM sideband tones, we designed them to have relatively narrow bandwidths of 20 MHz bandwidth for mode 1 with 7.3 MHz bandwidth for mode 2. The signals are further amplified with two cryogenic HEMT amplifiers, model number LNF4\_8. Due to a high noise temperature of HEMT 1, we added a Josephson traveling wave parametric amplifier\cite{Macklin2015Sup} immediately before it.

In order to route signal from mode $i$ only to the corresponding LJPA, the pin coupling to mode 1 is placed at the node of mode 2 (see Fig. \ref{fig:Setup}), so that photons from mode 2 do not couple to it. Placement of pin 2 and the choice of $\kappa_1>\kappa_2$ reduces leakage of mode 1 signal into pin 2. A small amount of signal from modes 1 and 2 is also lost through the input pin. We estimate the total reduction in quantum efficiency due to these loss channels to be approximately $10\%$ for each mode. The total quantum efficiencies are $\eta_1=0.41$ and $\eta_2=0.49$ for modes 1 and 2 respectively.

\setcounter{figure}{0}
\renewcommand{\figurename}{\textcolor{black}{\textbf{Extended Data\,Fig}}}

\begin{figure}
{\includegraphics[width=0.9\linewidth, bb=0 0 350 300]{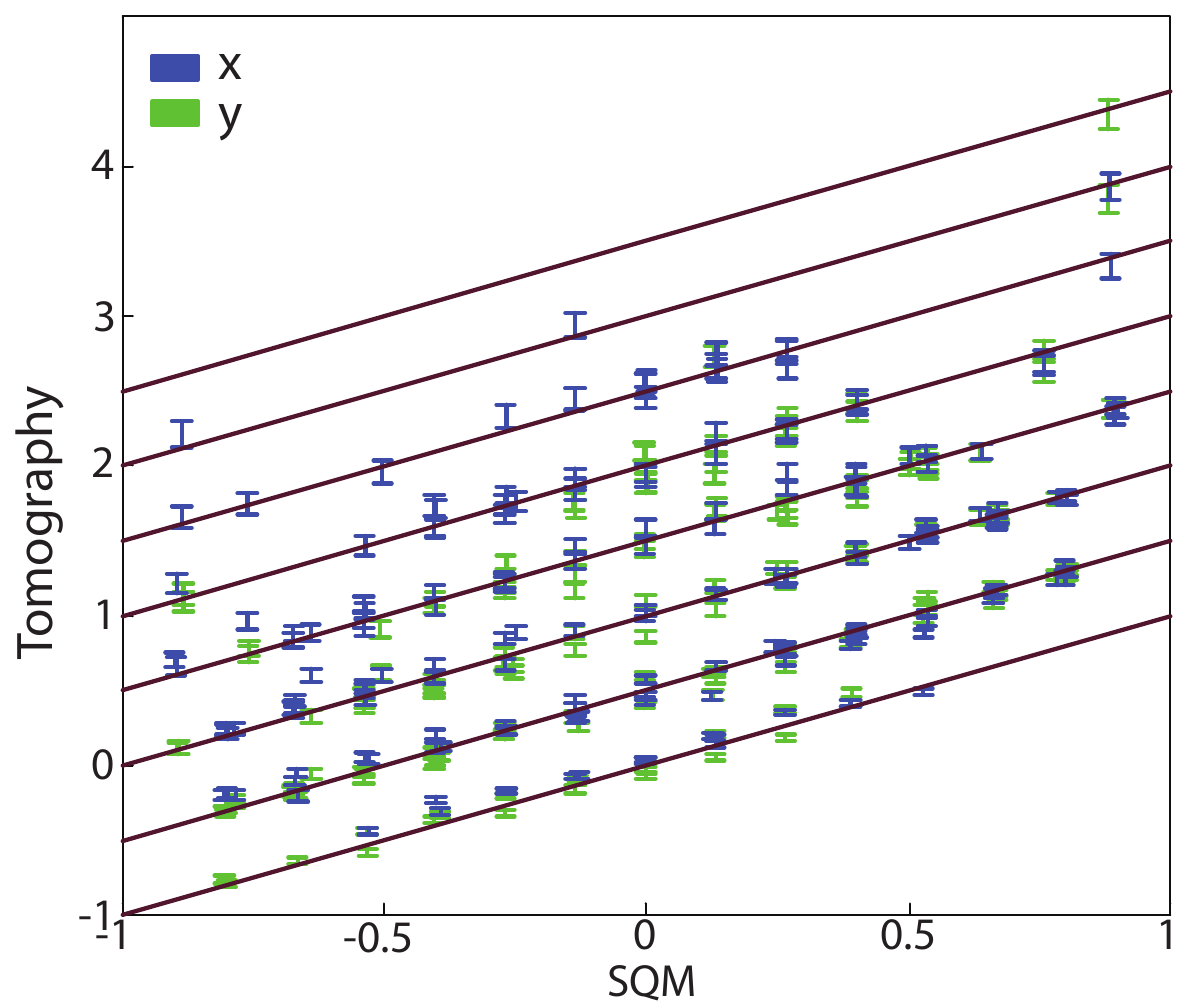}}
\caption{\textbf{SQM vs tomographic validation} Tomographic validation of a set of trajectories initialized in the state $y=-1$ and tracked for $3 \mu \mathrm{s}$, with the axes being perpendicular. Horizontal axis indicates Bloch sphere coordinate as predicted by trajectory reconstruction. Vertical axis represents the coordinate reconstructed from post-selected tomography data. Error bars are derived from the Poison statistics of the qubit measurement. Data are staggered vertically to allow for clearer visualization by adding one of seven arbitrary offsets. These offsets are chosen according to the number of measurements contributing to each data point.}
\label{fig:SuppTomoValid}
\end{figure}

\textbf{Measurement Calibration.} We generate SQM sideband tones by mixing a common local oscillator (LO) set to the corresponding cavity frequency with 40 MHz tones and controllable DC offsets input to the I and Q ports of a mixer. The output is followed by an amplifier model ZRON8G+ and then passed through a home built notch filter which attenuates the carrier by $\sim 25$ dB relative to the sideband tones. The filter and DC offset controls allow suppression of the LO to the $10^{-4}$ photon number level, which eliminates spurious rotations about the measurement axis. We balance the relative amplitudes of the sideband tones by measuring the induced cooling or heating with Ramsey interferometry\cite{Murch2012}, and then adjusting the relative phase between the I and Q 40 MHz signals. 

Since the sideband tones stark shift the qubit, all qubit drives and pulses are applied with the sidebands present and at the Stark-shifted qubit frequency, so that the experiment takes place entirely in this frame. The Rabi frequency is chosen such that the measurement operates in the unresolved sideband regime $\Omega_R \gg \kappa$. The sideband tone powers are set such that the measurement rate is much slower than the cavity mode bandwidths, so that the polaron transformation\cite{Gambetta2008} holds when both measurements are on.

Temperature drifts in the room-temperature electronics lead to instability of the Rabi frequency, which induces dephasing along the $\sigma_z$ axis of the qubit in the measurement frame. To efficiently and precisely measure this drift, we perform a $1~\mu s$ SQM followed by a lab-frame readout of $\sigma_z$. We interleave these measurements between the taking of trajectory data. This duration is chosen so that $\langle \sigma_z\rangle \approx 0$, which means that the readout is maximally sensitive to drift in the Rabi frequency. We use this signal to stabilize the Rabi frequency by optimizing the power of the Rabi drive every few seconds.

\begin{figure}
{\includegraphics[width=0.85\linewidth, bb=0 0 350 200]{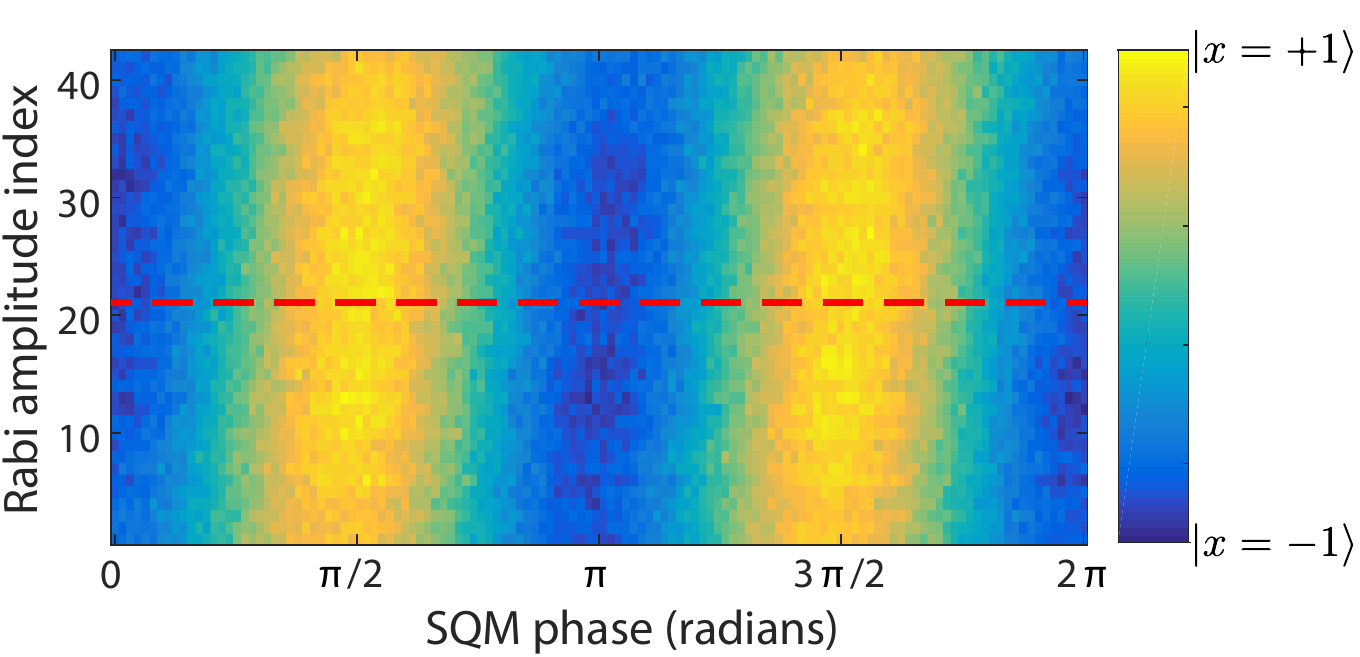}}
\caption{\textbf{SQM calibration.} Integrated SQM signal for measurement calibration. Color axis is scaled to the outcomes in which the qubit is aligned or anti-aligned to the measurement axis. Vertical sweep is performed by sweeping the modulation amplitude to drive approximately $40~\text{MHz}$ Rabi oscillations. The red dashed line indicates the index at which the Rabi frequency was determined to be $40~\text{MHz}$.}
\label{fig:SuppMeth1}
\end{figure}

After setting up the SQM sidebands, one must carefully set the feedback so that it stabilizes the Rabi frequency to 40 MHz. Furthermore, as the Rabi drive ramps up to its maximal power over a finite duration, 
the relative angle that the initial qubit state makes with the SQM axis is also unknown and must be measured.
We jointly calibrate these parameters by sweeping both the SQM axis and the Rabi frequency. We perform this calibration by measuring 
the SQM signal integrated over $7~\mu s$ as a function of these two sweep parameters, as shown in extended data Fig. \ref{fig:SuppMeth1}. When the initial qubit state is aligned or anti-aligned with the measurement axis and the Rabi frequency is precisely 40.00 MHz, then the qubit remains aligned throughout the duration of the measurement, and the integrated output signal takes its maximal or minimal value. Thus the Rabi frequency is 40.00 MHz when this contrast is maximal. We have indicated this region of the plot with a red line. Rotation of the qubit during the measurement leads to the slight tilt evident in extended data Fig. \ref{fig:SuppMeth1}, which means that a 1D sweep of just the phase or Rabi amplitude would be insufficient to accurately find the global maximum.
Due to instability of the Rabi frequency without feedback, the units in the vertical axis of the plot are not precisely known. While taking these data, we also perform the measurements described in the previous paragraph for feedback stabilization of the Rabi frequency and record the measurement outcome associated with each setting of the Rabi frequency. Once the desired Rabi frequency is identified, we look up the associated feedback measurement outcome and use this for value for stabilization.

In order to measure quantum trajectories, we require an accurate estimate of the SQM rates $\Gamma_{i}$ and quantum efficiencies $\eta_i$. The former we measure by preparing $|+\rangle$ and then making a Ramsey measurement. We measure the quantum efficiency by preparing states aligned and anti-aligned with the SQM. Histograms of the integrated measurement records yield a pair of Gaussians which separate as a function of time. The quantum efficiency is given by\cite{Korotkov2011Sup}.

\begin{equation}
\eta_i = \frac{(\mu_\uparrow - \mu_\downarrow)^2}{8 \tau \sigma^2 \Gamma_{i} }
\end{equation}

where $\mu_{\uparrow/\downarrow}$ is the mean of the Gaussian for the aligned/anti-aligned state preparation, $\sigma$ is the average standard deviation of the Gaussians and $\tau$ is the measurement duration.

In order to validate our system calibration and quantum trajectory reconstructions, we apply one of seven tomography pulses $\{\text{Identity}, \pi/2_x, -\pi/2_x, \pi/2_y, -\pi/2_y, \pi_x, -\pi_x \}$ at the end of each trajectory followed by a dispersive readout in the lab frame. Extended data Fig. \ref{fig:SuppTomoValid} shows comparison between tomographic estimates of $\langle \sigma_x \rangle$ and $\langle \sigma_y \rangle$ and their corresponding estimates from trajectory reconstruction. To generate the tomographic validation data, we have partitioned the Bloch sphere into a $15\times 15\times 15$ grid for post-selection. Each data point is a tomographic measurement of all trajectories which were predicted to end within the corresponding voxel. 57\% of data points lie within error bars, indicating good agreement and low contribution of systematic errors.

\textbf{Single Quadrature Measurement.} Our system is described by the following dispersive Hamiltonian

\begin{align}
\label{eq:HTot}
& H = H_q + H_\text{drive} + \sum_{i=\{1,2\}} (\omega_{c,i} + \chi_i\sigma_z)a_i^\dagger a_i \\ \nonumber
& H_q = \frac{\omega_q}{2} \sigma_z + \Omega_R(\sigma + \sigma^\dagger)\cos(\omega_d t) \\ \nonumber
H_\text{drive} &= \sum_{i=\{1,2\}} \epsilon_i(t) a_i^\dagger + \epsilon_i(t)^* a_i
\end{align}

where $\epsilon_i(t) \in \mathbb{C}$
is the coherent drive that will represent the SQM sideband drives. 
The master equation for a cavity which is damped at rate $\kappa_i$ and monitored with quantum efficiency $\eta_i$ is

\begin{align}
\label{eq:CavSME}
d\rho = -i [H, \rho] dt 
+ \sum_{i=\{1,2\}} \kappa_i \mathcal{D}[a_i] \rho ~dt + \sqrt{\kappa_i \eta_i} \mathcal{H}[a_i e^{i \phi_i}] \rho ~dW_i
\end{align}

where $\mathcal{D}[X]\rho = X\rho X^\dagger - (X^\dagger X\rho + \rho X^\dagger X)/2$ is the dissipation superoperator, $\mathcal{H}[X]\rho = X\rho + \rho X^\dagger-\langle X\rho + \rho X^\dagger\rangle \rho$ and $\phi_i$ is the amplification axis of the phase sensitive amplifier used to monitor mode $i$. $dW_i$ is a Gaussian distributed variable with variance $dt$ extracted from measurement record $i$. For the derivation, it suffices to consider just one of the two cavity modes. To measure a single quadrature of the qubit, we drive the cavity with a pair of sidebands detuned by $\pm \Omega_R$ from the cavity resonance, and with phases 
$\pm (\delta-\text{atan}(\kappa/2\Omega_R))$. As $\Omega_R \gg \kappa$, the latter term can be ignored in our experiment. We choose this drive $\epsilon(t) = -i \bar{a}_0 \sqrt{\Omega_R^2 + \kappa^2/4}\sin(\Omega_R t + \delta - \text{atan}(\kappa/2\Omega_R))e^{-i \omega_c t}$ so that the cavity internal displacement due to the sidebands takes the form

\begin{equation}
\label{eq:SidebandDrive}
\bar{a}(t) = \bar{a}_0 \cos(\Omega_R t + \delta) e^{-i \omega_c t} ~ \rightarrow ~ \bar{a}_0 \cos(\Omega_R t+\delta).
\end{equation}

The right arrow indicates that we have transformed into the interaction picture with respect to the cavity. We take $\bar{a}_0$ to be real for simplicity and define the displaced cavity operator $d = a-\bar{a}$. We first transform the cavity dissipation terms

\begin{align}
\kappa \mathcal{D}[a]\rho &= \kappa \mathcal{D}[d + \bar{a}]\rho = \kappa \mathcal{D}[d]\rho - i[H_\text{dis}, \rho] \\ \nonumber
H_\text{dis} &= -\frac{i \kappa}{2}(\bar{a} d^\dagger - \bar{a}^* d) \\ \nonumber
\sqrt{\kappa \eta} \mathcal{H}[a e^{i \phi}] &= \sqrt{\kappa \eta} \mathcal{H}[d e^{i \phi}].
\end{align}

This change of variables is equivalent to transforming the Hamiltonian by the displacement operator $\exp(\bar{a}a^\dagger - \bar{a} a)$, and thus adds a term $H' = i(\dot{\bar{a}} d - \dot{\bar{a}} d^\dagger)$ to the Hamiltonian. Note that $H_\text{dis}$ and $H'$ cancel the drive term $H_\text{drive}$ from \erf{eq:HTot}. Writing the remaining terms of $H$ for one cavity mode in the interaction picture with respect to the cavity and in the frame of the qubit drive (\textit{i.e.} transforming the Hamiltonian by $\text{exp}[i \omega_c a^\dagger a t + i \omega_d \sigma_z t/2 ]$) and applying the rotating wave approximation, we find

\begin{align}
H_\text{int} &= \chi a^\dagger a \sigma_z + H_{q, \text{int}}\\ \nonumber
&= \chi[(d^\dagger + \bar{a}^*)(d+\bar{a})]\sigma_z  + H_{q, \text{int}} \\ \nonumber
&= \chi \Big{[}\bar{a}_0 \cos(\Omega_R t + \delta)(d^\dagger + d) + \frac{\bar{n}_0}{2} +\\ \nonumber
& ~~~~ \bar{n}_0\frac{\cos(2\Omega_R t + 2\delta)}{2} + d^\dagger d \Big{]} \sigma_z  + H_{q, \text{int}} \\ \nonumber
H_{q, \text{int}} &= \frac{1}{2}[(\omega_q-\omega_d)\sigma_z + \Omega_R \sigma_x]
\end{align}

where we have defined $\bar{n}_0 = \bar{a}_0^2$. Choosing the Rabi drive to be resonant (\textit{i.e.} $\omega_q -\omega_d = -\chi \bar{n}_0$), we diagonalize the qubit drive term of the Hamiltonian by going into the Hadamard frame ($\sigma_z \leftrightarrow \sigma_x$), and then going into a frame rotating at the Rabi frequency ($\text{exp}[i\Omega_R \sigma_z t/2]$). These transformations eliminate $H_{q, \text{int}}$ and map $\sigma_z$ to $\sigma e^{-i \Omega_R t} + \sigma^\dagger e^{i\Omega_R t}$, so the Hamiltonian becomes

\begin{align}
H_\text{q-frame} &= \frac{\chi \bar{a}_0}{2}(e^{i\Omega_R t + i \delta} + e^{-i \Omega_R t - i\delta}) \times\\ \nonumber 
&(d^\dagger + d)(\sigma e^{-i \Omega_R t} + \sigma^\dagger e^{i \Omega_R t})  \\ \nonumber
& + \chi \Big{[} \bar{n}_0 \frac{\cos(2\Omega_R t + 2\delta)}{2} + d^\dagger d \Big{]}(\sigma e^{-i \Omega t} + \sigma^\dagger e^{i\Omega_R t}).
\end{align}

Dropping terms which rotate at $\Omega_R$ or $2\Omega_R$, we are left with

\begin{align}
\label{eq:HEffSuppMeth}
H_\text{q-frame} &= \frac{\chi \bar{a}_0}{2}(d^\dagger + d)(\sigma e^{i\delta} + \sigma^\dagger e^{-i \delta}) = \tilde{g}\sigma_\delta(d^\dagger + d) \nonumber \\
\sigma_{\delta} &\equiv \cos(\delta) \sigma_x + \sin(\delta) \sigma_y \nonumber \\
\tilde{g} &\equiv \frac{\chi \bar{a}_0}{2}
\end{align}

which is a qubit-state-dependent cavity drive Hamiltonian. 

If in addition to the sideband tones, a small amount of LO leakage (photons at the frequency of the mode) is also present at the cavity input, then Eq. \ref{eq:SidebandDrive} would read $\bar{\alpha}_0 \cos(\Omega_R t+\delta) + \bar{a}_\text{LO}$, and the effective Hamiltonian would instead be

\begin{align}
H_\text{q-frame} = \tilde{g}\sigma_\delta(d^\dagger + d + 2\text{Re}[\bar{a}_\text{LO}]).
\end{align}

Thus LO leakage induces unwanted rotations about the $\sigma_\delta$ axis. 

If the cavity starts in the ground state when this Hamiltonian is turned on, then it remains in a coherent state at all times. To derive an effective master equation for the qubit, we must eliminate the cavity entirely from \erf{eq:CavSME}. This may be accomplished by first applying a transformation which displaces the cavity to its ground state, which is known as the polaron transformation\cite{Gambetta2008}

\begin{align}
\label{eq:EOM}
U &= |e_\delta\rangle\langle e_\delta| D[\alpha_{e}] + |g_\delta\rangle\langle g_\delta| D[\alpha_{g}] \\ \nonumber
\dot{\alpha}_{e/g} &= -i \epsilon_{e/g} - \frac{\kappa}{2} \alpha_{e/g} ~~~~~ \epsilon_{e/g} = \pm\tilde{g}
\end{align}

where $D$ is the cavity displacement operator, $|e_\delta/g_\delta \rangle$ are the eigenstates of $\sigma_\delta$ and $\alpha_{e/g}$ is the cavity displacement conditional on the qubit state. Because the second line is the classical equation of motion for a resonantly driven cavity with drive rate $\epsilon$ and damping $\kappa$, this transformation maps the cavity to its ground state at all times, allowing us to trace it out. Transforming back from the polaron frame to the qubit frame, we find that the effective stochastic master equation for the qubit is

\begin{align}
\label{eq:RTerms}
d\rho &= \mathcal{L} \rho ~dt + \frac{\sqrt{\Gamma_{M}}}{2} \mathcal{H}[\sigma_\delta]\rho ~dW - i \frac{\sqrt{\Gamma_{M}'}}{2}[\sigma_\delta, \rho] dW \\ \nonumber
\mathcal{L}\rho &= -i\frac{B}{2}[\sigma_\delta, \rho] + \frac{\Gamma}{2} \mathcal{D}[\sigma_\delta] \rho \\ \nonumber
B &= 2 \tilde{g} \text{Re}(\alpha_e+\alpha_g) ~~~~~ \Gamma = -2 \tilde{g} \text{Im}(\alpha_e - \alpha_g) \\ \nonumber
\sqrt{\Gamma_{M}} &= \sqrt{\kappa \eta} |\beta| \cos(\phi - \theta_\beta) ~~~~~ \sqrt{\Gamma_{M}'} = \sqrt{\kappa \eta} |\beta| \sin(\phi-\theta_\beta) ~~~~~ \\ \nonumber &\beta = \alpha_e-\alpha_g
\end{align}

where $\theta_\beta = \text{arg}(\beta)$ is the angle of the cavity displacement axis $\beta$ in the IQ plane and $\Gamma$ is the dephasing rate. Substituting the equations of motion for the cavity (Eq.\ref{eq:EOM}) into the remaining expressions of Eq.\ref{eq:RTerms}, we find

\begin{align}
\label{eq:SME}
d\rho &= \frac{4 \tilde{g}^2}{\kappa}(1-e^{-\kappa t/2}) \mathcal{D}[\sigma_\delta]\rho ~dt ~+  \\ \nonumber
&~~\sqrt{\frac{4 \tilde{g}^2}{\kappa} \eta}(1-e^{-\kappa t/2}) \mathcal{H}[e^{i \phi} \sigma_\delta]\rho ~dW \\ \nonumber
 &\implies \Gamma = \frac{8 \tilde{g}^2}{\kappa} = \frac{2\chi^2 \bar{n}_0}{\kappa}.
\end{align}

We align the LJPA amplification axis to the axis of displacement arising in Eq. \ref{eq:HEffSuppMeth}, so that $\phi = \theta_\beta$. Application of sidebands to another mode of the cavity does not change the above derivation except to add an additional measurement also modeled by Eq. \ref{eq:SME}. Neglecting time-scales of order $1/\kappa$, measurement of two observables $\sigma_{\delta_1}$ and $\sigma_{\delta_2}$ is modeled by

\begin{align}
\label{eq:SMEFinal}
d\rho = \sum_{i=\{1,2\}} \frac{\Gamma_i}{2} \mathcal{D}[\sigma_{\delta_i}]\rho ~dt + \sqrt{\frac{\Gamma_i \eta_i}{2}}\mathcal{H}[\sigma_{\delta_i}]\rho~dW \\ \nonumber
V_i dt = \langle \sigma_{\delta_i} \rangle dt + \frac{dW_i}{\sqrt{2 \eta_i \Gamma_i}}
\end{align}

where $V_i$ is the measurement signal at time t, normalized appropriately. Equation \ref{eq:SMEFinal} can also be converted to a Fokker Planck equation, which propagates probability distributions of $\rho$. We use the latter to generate theory plots for Fig. \ref{fig:FokPlanck}. This stochastic master equation is generated by the following measurement operator

\begin{align}
\label{eq:XYMeasOpp}
\Omega(V) &= \exp \left[\sum_{i={1,2}} -\frac{\Gamma_{i} \eta_i}{2} \left(V_i(t)-\sigma_{\delta,i}\right)^2 dt \right] \\ \nonumber
\rho(t+dt) &= \mathcal{E}_{1-\eta_i}\frac{\Omega \rho(t) \Omega^\dagger}{\text{Tr}[\Omega \rho(t) \Omega^\dagger]} \\ \nonumber
\end{align}

where $\mathcal{E}_{1-\eta_i}$ is a superoperator which models dephasing due to finite quantum efficiency. To assure positivity of the state when $dt$ is taken to be finite, we use Eq. \ref{eq:XYMeasOpp} to numerically propagate quantum trajectories and also to calculate probabilities for maximum likelihood reconstruction.


\bibliographystyle{natureVV2}

\newpage



\end{document}